\documentstyle[epsfig]{article}
\begin{document}
\begin{titlepage}
\begin{flushright}
FERMILAB-Conf-97/240-T\\
\end{flushright}
\vspace{1cm}
\begin{center}
{\Large\bf Extracting $\alpha_S$ from TEVATRON data
\footnote{Talk given at the ``5th International Workshop on 
          Deep Inelastic Scattering and QCD'', Chicago, April 1997.}\\}
\vspace{1cm}
{\large
Walter T. Giele\\}
\vspace{0.5cm}
{\it
Fermi National Accelerator Laboratory, P.~O.~Box 500,\\
Batavia, IL 60510, U.S.A.} \\
\vspace{0.5cm}
{\large July 1997}
\vspace{0.5cm}
\end{center}
\begin{abstract}
In this contribution we explore one of the many
possibilities of determining 
the strong coupling constant $\alpha_S$ at hadron colliders.
The method considered is quite
unique compared to other methods
in that the 
value of $\alpha_S$ is determined by the
``evolution rate'' of the parton density functions
rather than by the ``event rate''.
\end{abstract}
\end{titlepage}
\section*{Introduction and Motivation}

Hadron colliders will supply an increasing amount of data
with the upcoming high luminosity TEVATRON run and the
LHC project. Methods for extracting $\alpha_S$ and 
parton density functions from these data sets 
can therefore expect a steady improvement in the precision
over the coming decades.

Often it is claimed that hadron colliders cannot do such
precision measurements. A few comments can be made in
answer to this. First of all, a hadron collider measures
the value of $\alpha_S$ at many different values of the (partonic)
center of mass energy. This is in contrast to
$e^+e^-$-colliders where the center of mass energy is fixed.
In fact, at the TEVATRON the partonic center of mass energy 
useful for the $\alpha_S$-extraction
can go as high as 1 TeV (and at the LHC this will increase
by an order of magnitude). 
Secondly, a hadron collider can make accurate
measurements by selecting
appropriate observables. For this purpose we select in this
talk the {\it normalized} one jet inclusive transverse
energy distribution. The value of $\alpha_S$ will be determined
from the shape of the distribution. The major factor determining
the shape is the fraction
of quarks in the colliding hadrons. Due to the
evolution of the parton density functions 
the quark fraction at moderate parton fractions decreases
as the jet energy increases. 
This depletion is controlled by the strength
of the strong coupling constant. 

Because we look at a normalized quantity both the theoretical
(renormalization scale dependence) and experimental uncertainties
(e.g. luminosity uncertainty) are much smaller than one would expect.

\section*{Measurement and Methodology}
\begin{figure}[t!] 
\centerline{\epsfig{file=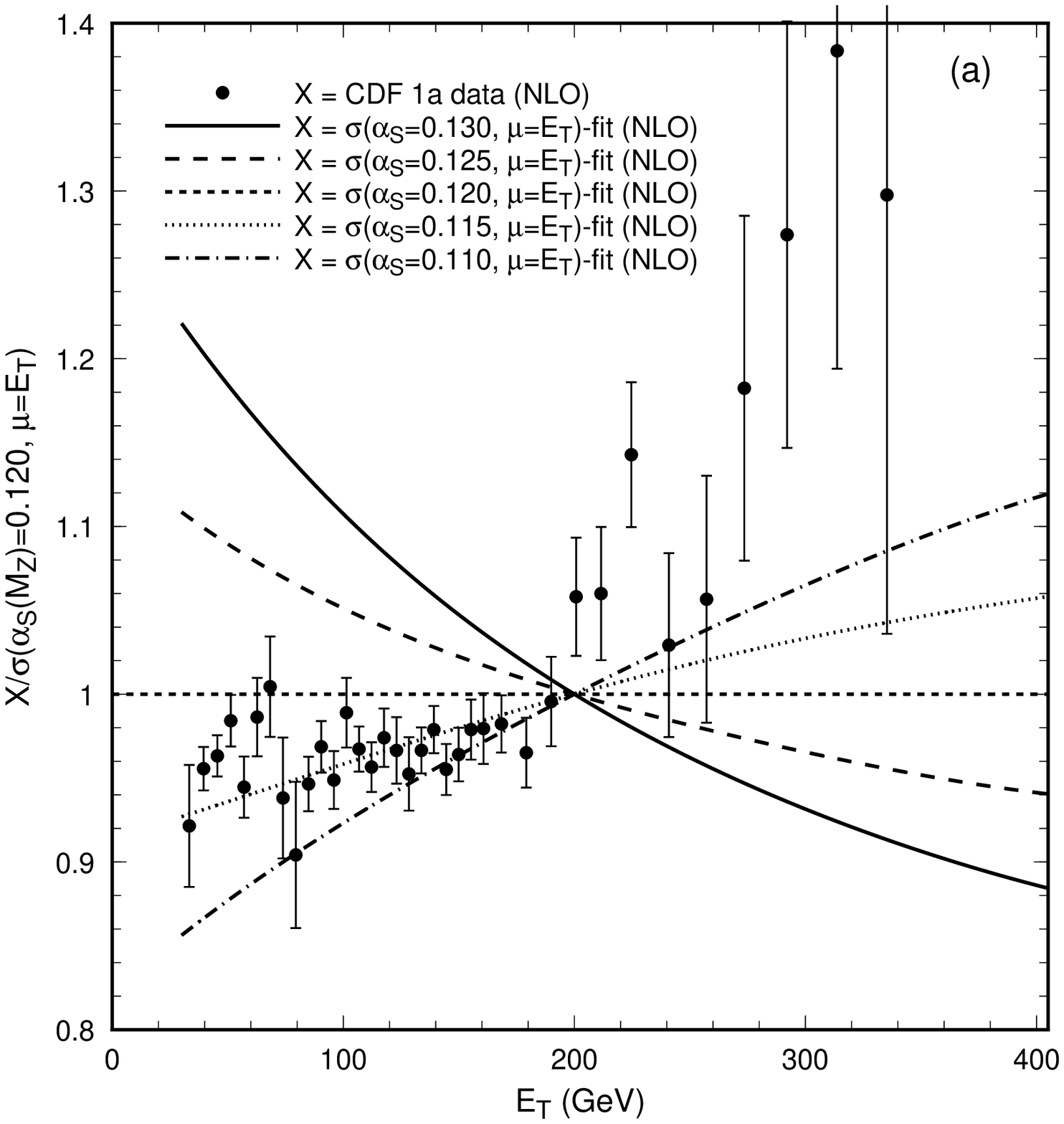,height=2.5in,width=2.5in}
            \epsfig{file=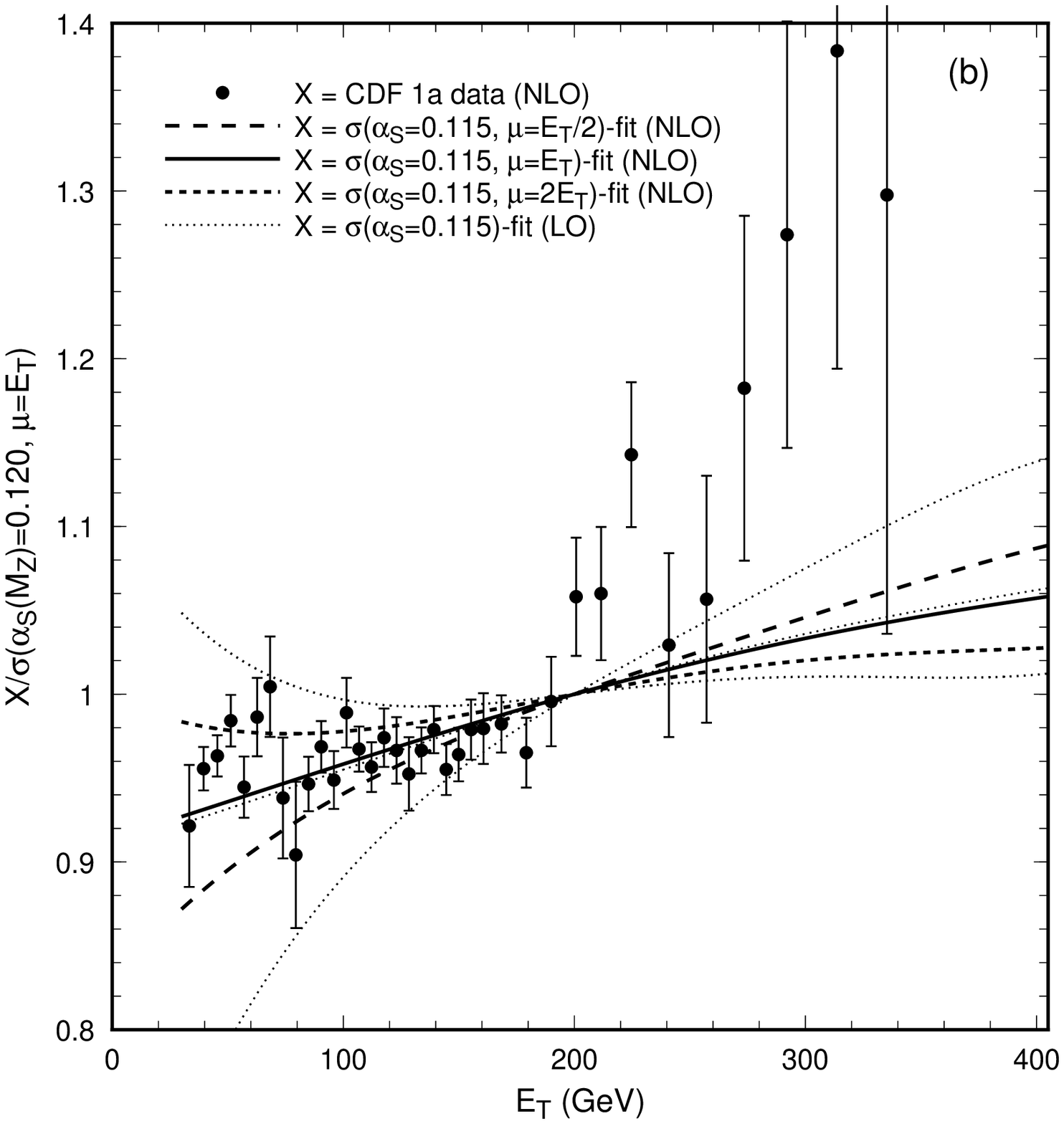,height=2.5in,width=2.5in}}
\caption[]{(a) The sensitivity of the distribution
to the value of $\alpha_S(M_Z)$. For comparison the CDF data 
\cite{CDF} is also shown. (b) The scale dependence
of both the leading (LO) and next-to-leading (NLO) predictions.}
\vspace*{10pt}
\end{figure}

As mentioned in the introduction, the observable used is
the normalized one jet inclusive transverse energy distribution.
As an example we use the published run 1a 
results from the CDF collaboration \cite{CDF}. For the theoretical
prediction we use the JETRAD monte carlo \cite{jetrad} with the cuts
and jet algorithm as close as possible to the experimental setup. The
MRSA' parton density functions \cite{MRS}, 
which allows varying $\alpha_S$, were used.
The renormalization/factorization scale, $\mu$, was 
chosen to be a constant, $\lambda$,
times the maximum jet transverse energy, $E_T$, in the event. Both data
and theory are divided by the ``reference'' theory prediction
which is given by: 
$\alpha_S(M_Z)=0.120$, $\mu=E_T$. To normalize
the distribution we choose the ratio to be equal
to unity at $E_T=200$ GeV . 

We show the $\alpha_S$-dependence in fig. 1a  and the scale
dependence in fig 1b. As can be seen the dependence on $\alpha_S$
is quite substantial compared to both the experimental and theoretical
uncertainties. Also note that the leading order (LO) and next-to-leading
order (NLO) results are quite close. The only difference between the two
predictions is the (expected) reduced scale dependence at NLO.
The method to extract $\alpha_S$ is now quite simple:
we minimize the $\chi^2$ to fit the theory to the data 
in fig. 1 between 30 GeV $\leq E_T\leq$ 200 GeV by varying both
$\alpha_S$ and $\lambda$. The results and, more importantly, the 
interpretation are discussed in the next section. Note that we do not
consider the systematic uncertainties at this point. They can be easily
included in the $\chi^2$-fit by building up the correlation matrix out
of the systematic uncertainties \cite{correlated}. 
Their inclusion is better left to 
the experimenters. Here we want to concentrate on the methodology.

\section*{Interpretation and Results}
\begin{figure}[t!] 
\centerline{\epsfig{file=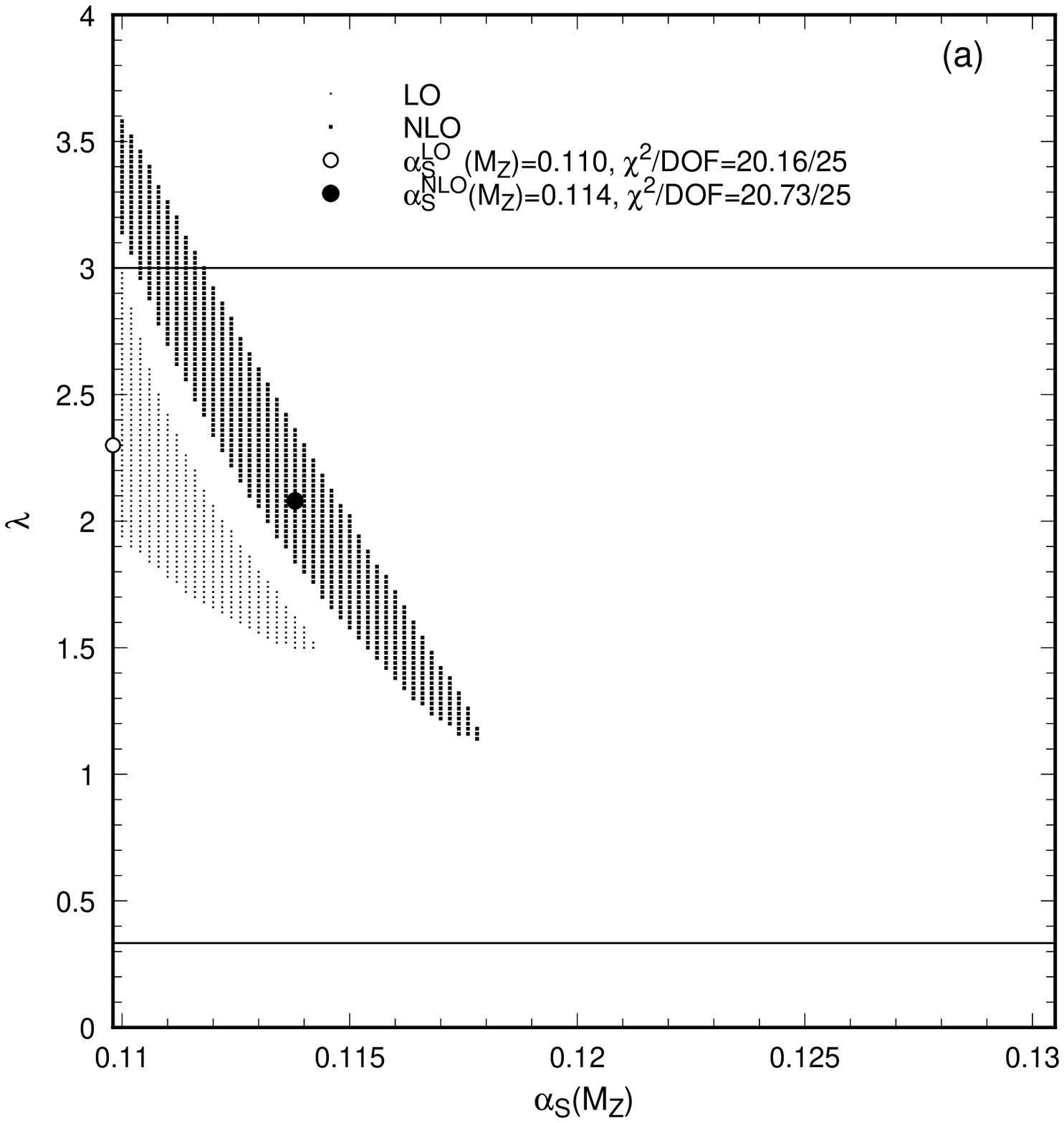,height=2.5in,width=2.5in}
            \epsfig{file=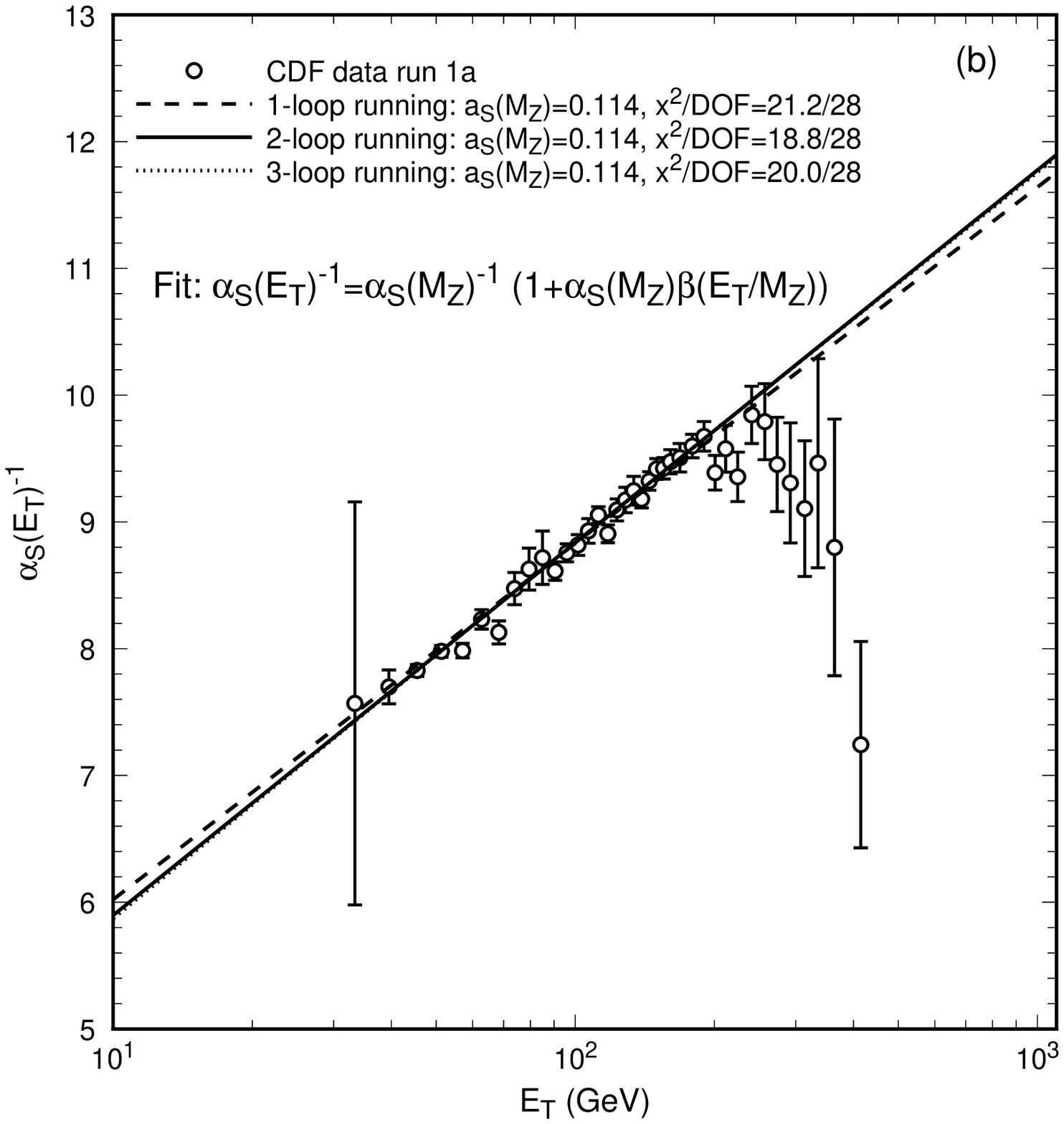,height=2.5in,width=2.5in}}
\caption[]{(a) The result of the $\chi^2$-fit to the data
for both the LO and NLO predictions as the minimum and the 1-$\sigma$ 
uncertainty ellipse. (b) A comparison of the extracted
central value of $\alpha_S^{NLO}(M_Z)$ with its one- two- and three-loop
evolution compared to $\alpha_S(E_T)$ extracted from the data using
the method of ref. \cite{GGY}.}
\vspace*{10pt}
\label{fig2}
\end{figure}
In fig. 2a we show the results of the minimalization 
procedure to fit to
the data. Both the minimum and the 1-$\sigma$ uncertainty ellipse is shown.
The figure contains all the information we can extract from the data. While
the central value is quite trivial to determine, the interpretation of the
uncertainty is not. The perfect answer (that is no 
renormalization scale uncertainty) would be a vertical strip. The uncertainty
would then simply be the width of the strip
independent of the choice of $\lambda$. However, in fixed order
perturbative QCD we have a residual scale sensitivity due to the truncation
of the series. This is reflected in the slope of the ellipse-axis.
In fact
one could argue that the slope is 
the correct  measure of the theoretical uncertainty.
When comparing results from different experiments this slope could be used
to weight different experiments on their theoretical uncertainty. 
This all implies that the parameter $\lambda$ cannot be considered 
a fitting parameter as $\alpha_S$ is, nor is it in a direct manner 
related to the theoretical uncertainty.
Note that when the data accuracy increases (e.g. the CDF/D0  run 1b data)
the fit of the theory to the data will become more strained and 
the variation of $\lambda$ will become more constrained.
This does not indicate that the theoretical uncertainty is decreasing.
On the contrary, this means that the NLO 
prediction is becoming more and more inadequate to describe 
the data and even higher order calculations are needed.

For the moment we use a naive procedure 
to quote the theoretical uncertainty.
The experimental uncertainty is taken to be the width of the ellipse at the 
minimum, while the theoretical uncertainty is taken as the variation within
1-$\sigma$ for scales between $1/3 \leq \lambda\leq 3$.
The results are
\begin{eqnarray}\nonumber
\alpha_S^{LO}(M_Z)&=&0.110\pm 0.001\mbox{(stat)}\pm 0.004\mbox{(theory)}\\
\alpha_S^{NLO}(M_Z)&=&0.114\pm 0.001\mbox{(stat)}\pm 0.004\mbox{(theory)}\ .
\nonumber \end{eqnarray}
Alternatively, one could argue that the difference between the LO and NLO
value of $\alpha_S$ should be larger than the difference between the NLO and
NNLO value of $\alpha_S$, giving an alternative, but equal, estimate on
the theoretical uncertainty of $0.004$.

\section*{Conclusions and Outlook}
We have used the normalized one jet inclusive transverse energy distribution
to extract $\alpha_S(M_Z)$. The fact that we used the normalized distribution
reduces the experimental and theoretical uncertainty significantly. 
To improve the results we need to fit the parton density functions 
(specifically the gluon) together with the value of $\alpha_S$. 
This should remedy the obvious discrepancies between the data and theory
for  $E_T > 200$ GeV (see ref. \cite{WuKi}). Such a measurement would not
only give us $\alpha_S$, but simultaneously a 
{\it true} NLO determination of the
gluon parton density function.
In fig. 2b we finally show, as a cross check, the comparison between the
$\alpha_S$ extracted in this talk and the 
$\alpha_S$-values determined in each $E_T$-bin, obtained
using the methods of ref. \cite{GGY} (incorporating parton 
density functions with varying $\alpha_S$). As can be seen
the agreement between the two methods is quite good. 

The new run 1b data from the CDF and D0 collaboration
will severely test the NLO description of the data and higher
order calculations might be needed to describe these results
and extract the gluon parton density function and $\alpha_S$.


\begin{thebibliography}{9}
\bibitem{CDF} F.~Abe et al., {\it Phys. Rev. Lett.} {\bf 77} 438 (1996).
\bibitem{jetrad} W.~T.~Giele et al., {\it Nucl. Phys.} {\bf B403} 633 (1993).
\bibitem{MRS} A.~D.~Martin et al., {\it Phys. Lett.} {\bf B387} 41 (1996).
\bibitem{correlated} See e.g. S.~Alekhin, {\it hep-ph} 9611213 (1996).
\bibitem{WuKi} J.~Huston et al., {\it Phys. Rev. Lett.} {\bf 77} 444 (1996).
\bibitem{GGY} W.~T.~Giele et al., {\it Phys. Rev.} {\bf D53} 120 (1996). 
\end{thebibliography}
\end{document}